\documentclass[11pt]{article}
\usepackage{graphicx}
\newcommand{\BABARPubYear}    {04}

\newcommand{\BABARConfNumber} {48}
\newcommand{\SLACPubNumber} {10623}
\newcommand{\LANLNumber} {0000}

\input babarsym

\long\def\inst#1{\par\nobreak\kern 4pt\nobreak
    {\it #1}\par\vskip 10pt plus 3pt minus 3pt}

\newcommand{\lumi}    {110.7\invfb}

\def\kk2f       {\mbox{\tt KK2f}\xspace}

\def\sina    {\pt / E_{\rm missing} }

\def\nelectrons{9668}
\def\nmuons{6201}
\def\febkgd{0.210\pm 0.016}

\def\effe{0.0455\pm 0.0004}
\def\effm{0.0291\pm 0.0003}

\def\bre{8.52\pm 0.09\pm 0.40}
\def\brm{8.54\pm 0.11\pm 0.45}

\def\taux       {\taum \rightarrow  X^- \nut }

\def\taufivezero {\taum \rightarrow  3h^- 2h^+  \nut }
\def\taufiveone  {\taum \rightarrow  3h^- 2h^+  \pi^0 \nut }

\def\taupkk     {\taum \rightarrow  \pim \KS  \KS   \nut } 
\def\tauhhhk    {\taum \rightarrow  h^-h^-h^+ \Kz   \nut }

\begin{document}
{\pagestyle{empty}

\begin{flushright}
\babar-CONF-\BABARPubYear/\BABARConfNumber \\
SLAC-PUB-\SLACPubNumber \\
hep-ex/\LANLNumber \\
July 2004 \\
\end{flushright}

\par\vskip 5cm

\begin{center}
\Large \bf Study of the $\taufivezero$ Decay 
\end{center}
\bigskip

\begin{center}
\large The \babar\ Collaboration\\
\mbox{ }\\
\today
\end{center}
\bigskip \bigskip

\begin{center}
\large \bf Abstract
\end{center}
A preliminary measurement of the branching fraction of 
the $\taufivezero$ decay ($h= \pi, \kaon$) with 
the \babar\ detector is found to be $(\bre) \times 10^{-4}$, where the
first error is statistical and the second is systematic.
The data show evidence that the $\rho$ resonance plays
a strong role in the decay of the \mtau lepton to five charged hadrons.
\vfill
\begin{center}

Submitted to the 32$^{\rm nd}$ International Conference on High-Energy Physics, ICHEP 04,\\
16 August---22 August 2004, Beijing, China

\end{center}

\vspace{1.0cm}
\begin{center}
{\em Stanford Linear Accelerator Center, Stanford University, 
Stanford, CA 94309} \\ \vspace{0.1cm}\hrule\vspace{0.1cm}
Work supported in part by Department of Energy contract DE-AC03-76SF00515.
\end{center}

\newpage
\begin{center}
\small

The \babar\ Collaboration,
\bigskip

%
B.~Aubert,
R.~Barate,
D.~Boutigny,
F.~Couderc,
J.-M.~Gaillard,
A.~Hicheur,
Y.~Karyotakis,
J.~P.~Lees,
V.~Tisserand,
A.~Zghiche
\inst{Laboratoire de Physique des Particules, F-74941 Annecy-le-Vieux, France }
% ...

\end{center}\newpage

} 

The semi-leptonic decays of the \mtau lepton are an ideal area
for studying strong interaction effects
(for example, see Ref.~\cite{stahl}).
The decay mode $\taux$ probes the matrix element of the left-handed
current between the vacuum and the hadronic state $X^-$ \cite{pich}.
Most of these studies have involved the decay of the \mtau to one or three
charged particles and any number of \piz mesons whereas decays of the \mtau 
to five charged particles have been limited by the small number of 
observed events \cite{PDG}.
This paper presents a preliminary measurement of the $\taufivezero$ decay 
($h= \pi, \kaon$) from a sample of over 15,000 such 
decays.\footnote{Charge conjugation is assumed throughout this paper.
In addition, a five charged particle state is not considered a
$\taufivezero$ decay if it was the result of a \KS decay.
No attempt has been made in this work to distinguish charged pions and kaons.}

This analysis is based on data recorded 
by the \babar\ detector at the \pep2\ asymmetric-energy \epem\ 
storage ring operated at the Stanford Linear Accelerator Center.
The data sample consists of \lumi\ recorded at
center-of-mass energy ($\sqrt{s}$)
of 10.58 \gev and 10.54 \gev between 1999 and 2003.
With an expected cross section for \mtau-pair production  
at the luminosity-weighted 
$\sqrt{s}$ of $\sigma_{\tau\tau} = (0.89\pm0.02)$ nb \cite{kk},
this data sample contains approximately 200 million \mtau decays.
Monte Carlo simulation is used to evaluate the 
background contamination and selection efficiency.
The \mtau pair events are simulated with the KK2f Monte Carlo event
generator \cite{kk} and the \mtau decays were modeled 
with Tauola \cite{tauola} according to measured rates \cite{PDG}.
 
The \babar\ detector is described in detail in \cite{detector}.
Charged particle  momenta are measured with a five-layer
double-sided silicon vertex tracker and a 40-layer drift chamber 
inside a 1.5-T superconducting solenoidal magnet.
A calorimeter consisting of 6580 CsI(Tl) 
crystals is used to measure electromagnetic-shower energy,
a ring-imaging Cherenkov detector is used to identify
charged hadrons, 
and an instrumented magnetic flux return (IFR) is used to
identify muons.

The \mtau pairs are produced back-to-back in the \epem CM frame.  
As a result it is convenient to divide the event into two hemispheres, 
each containing the decay products of a single \mtau lepton.
The analysis procedure selects events with one track in one 
hemisphere (tag hemisphere) and five tracks in the other hemisphere
(signal hemisphere).
The track in the tag hemisphere is required to be an electron
or muon to reduce background from non-\mtau events.

The event is divided into two hemispheres in the CM frame
based on the plane perpendicular to the thrust axis.
The thrust is calculated with the use of the tracks in the event.
The number of tracks in each hemisphere is used to determine
the topology of the event.
The tracks are required to have a minimum transverse momentum
with respect to the beam of less than $0.1\gevc$, 
a distance of closest approach to the production point in the transverse 
plane to the beam axis (DOCA$_{XY}$) of less than 1.5 cm 
and the absolute value of the distance 
of closest approach in the $z$-plane of less than 10 cm.
  
The background from non-\mtau sources (in particular, Bhabha 
scattering and 
two-photon production) is reduced with the use of the 
magnitude of the thrust of the event and the ratio
$\sina$ where \pt is the transverse component of the vector sum
of the momenta of all the tracks in the event and $E_{\rm missing}$ is the 
missing energy in the event.  
The $\sina$ variable is very effective in reducing the background from
two-photon production which tend to have low \pt and high $E_{\rm missing}$.
The thrust ($T$) is required to be in the range between 0.92 and 0.99.
Events are retained if they satisfy the following criteria
\begin{eqnarray*}
(\sina > 0.3  \;\; \mbox{\rm and} && 0.92 < T < 0.93) \;\; \mbox{\rm or} \\
(\sina > 0.2  \;\; \mbox{\rm and} && 0.93 < T < 0.94) \;\; \mbox{\rm or}  \\
(\sina > 0.1  \;\; \mbox{\rm and} && 0.94 < T < 0.95).  
\end{eqnarray*}
There is no requirement on $\sina$ if the thrust is between 0.95 
and 0.99.

Further reduction of the non-\mtau background is made by requiring
that the track in the tag hemisphere be identified as an electron or a muon.
Electrons are identified with the use of
the ratio of calorimeter energy to 
track momentum $(E/p)$, the ionization  loss in the tracking system 
$(\dedx)$, and the shape of the shower in the calorimeter.
Muons are identified with hits in the IFR and small energy deposits in the
calorimeter.

The momentum of the lepton in the tag hemisphere in the center-of-mass frame
is required to be less than $4\gevc$ to reduce background from 
Bhabha scattering and dimuon events.
Residual background from multihadronic events is reduced by cutting on
the number of clusters in the electromagnetic calorimeter 
within the tag hemisphere and not associated to the lepton.
It is required that there be at most one cluster with 
energy between 0.05 and 1 \gev.

Additional criteria are applied to the five tracks in the signal hemisphere
to reduce background from photon conversions and secondary decays.
The event is rejected if any of the tracks in the signal hemisphere
is identified as an electron or if any pair of oppositely charged tracks is
consistent with originating from a photon conversion.
The reconstructed mass of the five tracks is required to be 
less than $1.8 \gevcc$. 

It is also required that there be no $\piz$ candidates in the 
signal hemisphere.
The \piz finding algorithm first searches for two clusters 
(each of at least 50 \mev) in the electromagnetic calorimeter 
that reconstructs to the \piz mass (0.115 - 0.150 \gevcc).
Any residual clusters are considered $\piz$ mesons if their
energy is greater than $0.5 \gev$ and they are not associated with
any tracks.

A total of $\nelectrons$ and $\nmuons$ events are selected when an electron
or muon are identified in the tag hemisphere, respectively.
The background fractions in the electron and the muon tag samples 
are $\febkgd$.\footnote{The total background in the electron
and muon tag samples are evaluated independently and by coincidence,
are identical.}
The efficiencies for selecting the lepton plus $\taufivezero$ 
events are $\effe$ and $\effm$  in the electron and muon samples, respectively,
where the quoted uncertainty is the Monte Carlo statistical 
error\footnote{The efficiency is defined to the ratio of the number
of selected lepton plus $\taufivezero$ events divided by the number
of $\tau$-pair events with a $\taufivezero$.  
The branching fraction of the lepton
is incorporated into the selection efficiency.}.

The background is estimated by Monte Carlo simulation and tested with 
a dataset where the particular background is enhanced.
The sources of background in the electron tag sample\footnote{An
event is categorized as signal or background by whether or not the
decay in the signal hemisphere passes the five-prong selection.
An event is considered a signal event if
the track in the tag hemisphere is mis-identified as an electron
and the decay in the signal hemisphere passes the five-prong selection.}
can be broken down into the following categories:
$\taufiveone$ decays (6.8\%),
\mtau decays with one or three tracks and at least one $\piz$ (5.7\%), 
\mtau decays with a $\KS$ (5.2\%),
multihadronic events (2.9\%, primarily $\ccbar$ events)
and a residual amount from other \mtau decays (0.3\%).
The relative uncertainties range between 10 and 20\% for each background and 
reflect the statistical precision of the data and
Monte Carlo samples used to evaluate the backgrounds.
The sources of background in the muon-tag sample 
are almost the same as for the electron-tag sample.

The $\taufiveone$ background is validated with the use of
the energy of the $\piz$
found in five-prong events that contain a single $\piz$ in the 
signal hemisphere.
The background from \mtau one- and three-prong decays with a \piz arises when
one of the photons converts to an \epem pair or 
a $\piz \rightarrow \epem \gamma$ decay.
Decays are removed if there is an identified photon conversion or if 
any tracks in the event is considered an electron candidate.

The background from \mtau decays with 
at least one $\KS$ ($\taupkk$ and $\tauhhhk$ 
decays) is determined by fitting the mass distribution of $\pip \pim$ pairs
to obtain an estimate of the number of $\KS$ mesons.
The background estimation takes into account that the $\taupkk$ 
decays often have two identified $\KS$ mesons in a single event.
The uncertainty in the background from \mtau decays with \KS mesons
was found to be approximately 20\% and includes contributions from the 
fit uncertainty and the branching ratios of the background decay modes.
In addition, checks were made to ensure that the \KS background
was from \mtau decays and not multihadronic events.

The background from multihadronic events was estimated by selecting
events where the reconstructed mass of the five tracks in the signal 
hemisphere is above the \mtau mass.  
In addition, events with more than one cluster in the electromagnetic
calorimeter in the tag hemisphere were used to measure the
multihadronic background.

\begin{figure}[!htb]
\begin{center}
\includegraphics[height=7cm]{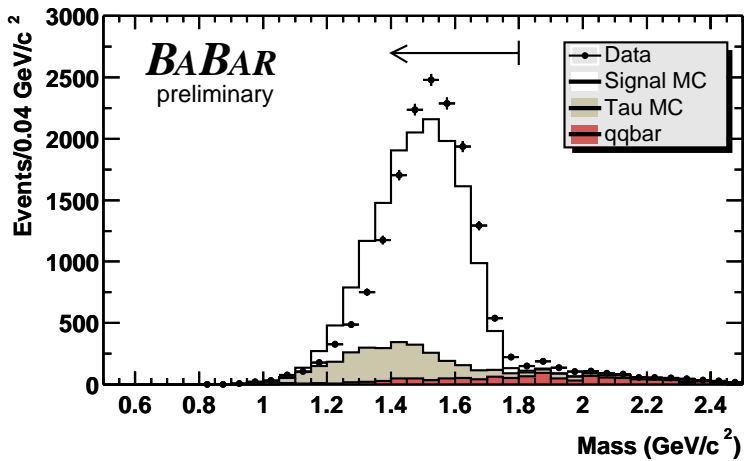}
\end{center}
\caption{
Reconstructed mass of the five tracks in the signal hemisphere 
after all other selection criteria are applied.
The points are the data and the histogram is the Monte Carlo
simulation for both the electron and muon tag samples, respectively.   
The unshaded and shaded histograms are 
the signal and background events.
The arrow indicates the selection requirement applied to the 
samples.
The Monte Carlo simulation is normalized to the luminosity of
the data sample.
}
\label{fig1}
\end{figure}

The branching fraction\footnote{The branching fraction is
defined as $B = \frac{N_{sel}}{2N}  \frac{1-f_{bkgd}}{\epsilon}$
where 
$N_{sel}$ is the number of selected events (1-prong lepton tag plus 
$\taufivezero$ candidate), 
$N$ is the number of tau pair events determined
from the cross section and luminosity,
$f_{bkgd}$ is the fraction of background, 
and $\epsilon$ is the efficiency for selecting 
lepton and $\taufivezero$ events.
}
of the $\taufivezero$ decay is found to be
$(\bre) \times 10^{-4}$ and $(\brm) \times 10^{-4}$ for the 
data selected by the electron and muon tags, respectively.
The first uncertainties is the statistical error and the second  systematic.

The systematic error includes contributions from 
the efficiency for reconstructing the six tracks in the event (3.1\%),
the luminosity and \tautau cross section (2.3\%), 
the $\piz$ finding algorithm (2.0\%),
the background in the five-prong sample (1.5\%), 
and the lepton identification in the tag hemisphere (1.0\%).
The numbers given above are for the electron tag sample.  
The systematic uncertainty for the muon tag data is slightly larger.
The systematic errors on the branching fraction determined with the
electron and muon samples are highly correlated and combining the
two results would result in no improvement in the total uncertainty.
As a consequence, the branching fraction obtained with the electron tag 
is used for the final result.

The error on the efficiency for selecting six tracks is based on studies 
of the efficiency for reconstructing a single track.
The error on the efficiency for reconstructing a track is estimated
to be 1.2\% for tracks with $\pxy < 0.3 \gevc$ and 0.5\% for tracks
with $\pxy > 0.3 \gevc$.
The errors were obtained from comparison of efficiencies of the
standalone track reconstruction in the silicon vertex tracker and the
drift chamber, and confirmed by the an independent analysis of the 
\mtau decays into three charged particles and neutrino. 
The variation of the
selection cuts such as the minimum transverse momentum of the track,
the number of tracks with hits in the silicon vertex tracker, 
and the sum of the DOCA$_{xy}$ of the five
tracks resulted in a negligible change in the branching fraction. 

The systematic error associated with the $\piz$-finding algorithm 
used to separate
$\taufivezero$ and $\taufiveone$ events was based on the $\piz$ 
energy distribution.
An excess of data over the Monte Carlo simulation was observed at 
low $\piz$ energy and a systematic uncertainty of 2\% 
in the selection efficiency was included to account for this discrepancy.

The systematic uncertainty for selecting electrons and muons in the tag 
hemisphere has been conservatively estimated to be 1\% and 2.5\% respectively.
Other consistency checks included varying the selection variables.
In addition, the selection efficiency was found to have no dependence on the 
reconstructed mass of the five tracks in the signal hemisphere.

In Fig.~\ref{fig1}, the mass of the five tracks in the signal hemisphere 
is presented.\footnote{All mass distributions shown in 
this paper are calculated assuming that the particles are pions.}
Tauola uses a phase space distribution for 
the $\taum \rightarrow  3\pim 2\pip  \nut$ 
decay \cite{tauola}.\footnote{Tauola 
does not generate any five-prong decays with charged kaons.}
The small samples of $\taufivezero$ decays recorded by other experiments
prior to this measurement find no disagreement with a phase-space distribution.
It is clear from Fig.~\ref{fig1} that a phase-space distribution does not 
give a good description of the mass of the five tracks.
This is not surprising as \mtau decays with two to four pions in the final 
state cannot be modeled with a phase-space distribution 
and one needs to include resonances.

\begin{figure}[!htb]
\begin{center}
\includegraphics[height=7cm]{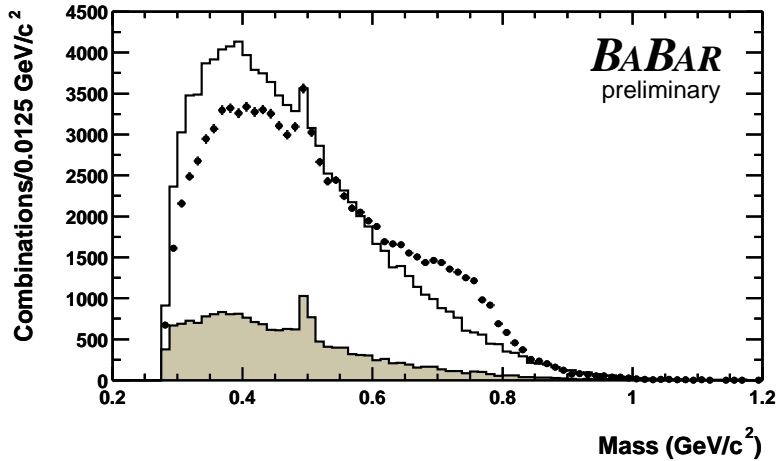}
\end{center}
\caption{
Reconstructed mass of $h^+ h^-$ pairs in the five tracks in the 
signal hemisphere.
The data are shown as points with error bars.
The unshaded and shaded histograms are the signal 
and background predicted by the Monte Carlo simulation.
The peak at $0.5 \gevcc$ are $\KS$ mesons which are not rejected 
by the selection.
}
\label{fig2}
\end{figure}

In three-prong \mtau decays clear evidence could be found for resonant
contributions by plotting the mass of $\pip \pim$ pairs (for example,
see Ref.~\cite{cleo-3prong}).
In Fig.~\ref{fig2} the mass of $h^+ h^-$ pair combinations is plotted
for all five-prong events in both the electron and muon tag samples. 
Evidence for the $\rho$ resonance at $0.77 \gevcc$ in the
$h^+ h^-$ mass distribution is apparent.
The observation of the $\rho$ in  $\taum \rightarrow  3\pim 2\pip  \nut$
decays is not unexpected.
There are three allowed isospin states for the 
$\taum \rightarrow  3\pim 2\pip  \nut$ decay mode (for example, 
see Ref.~\cite{sobie}) and two of these isospin states 
have the same quantum numbers as the $\rho$ resonance.

In summary, the \babar\ Collaboration has made a preliminary measurement 
of the $\taufivezero$ branching fraction,
$B(\taufivezero) = (\bre) \times 10^{-4}$.
The invariant mass distribution of $h^+ h^-$ pairs suggests
that the $\rho$ meson is 
abundantly produced in the $\taufivezero$ decay.

\hspace{0.25cm}

We are grateful for the 
extraordinary contributions of our \pep2\ colleagues in
achieving the excellent luminosity and machine conditions
that have made this work possible.
The success of this project also relies critically on the 
expertise and dedication of the computing organizations that 
support \babar.
The collaborating institutions wish to thank 
SLAC for its support and the kind hospitality extended to them. 
This work is supported by the
US Department of Energy
and National Science Foundation, the
Natural Sciences and Engineering Research Council (Canada),
Institute of High Energy Physics (China), the
Commissariat \`a l'Energie Atomique and
Institut National de Physique Nucl\'eaire et de Physique des Particules
(France), the
Bundesministerium f\"ur Bildung und Forschung and
Deutsche Forschungsgemeinschaft
(Germany), the
Istituto Nazionale di Fisica Nucleare (Italy),
the Foundation for Fundamental Research on Matter (The Netherlands),
the Research Council of Norway, the
Ministry of Science and Technology of the Russian Federation, and the
Particle Physics and Astronomy Research Council (United Kingdom). 
Individuals have received support from 
CONACyT (Mexico),
the A. P. Sloan Foundation, 
the Research Corporation,
and the Alexander von Humboldt Foundation.



\begin{thebibliography}{99}

\bibitem{stahl}
A. Stahl,
Physics with Tau Leptons,
Springer Tracts in Modern Physics,  Volume {\bf 160}  (2000). 

\bibitem{pich}
A. Pich,
Nucl.\ Phys.\ Proc.\ Suppl.\  {\bf 98}, 385 (2001).

\bibitem{PDG}
Particle Data Group, 
S. Eidelman {\it et al.}, Phys. Lett. {\bf B592}, 1 (2004).


\bibitem{kk}
B.~F.~Ward, S.~Jadach, and Z.~Was,
Nucl.\ Phys.\ Proc.\ Suppl.\  {\bf 116}, 73 (2003).

\bibitem{tauola}
S.~Jadach, Z.~Was, R.~Decker, and J.~H.~Kuhn,
Comput.\ Phys.\ Commun.\  {\bf 76}, 361 (1993).

\bibitem{detector}
\babar\ Collaboration,  B.~Aubert {\it et al.}, \nima{479}, 1 (2002).

\bibitem{cleo-3prong}
CLEO Collaboration, R.A. Briere {\it et al.},
\jprlBase\  {\bf 90}, 181802 (2003).

\bibitem{sobie}
R.J.Sobie, \jprBase\ {\bf D60} 017301 (1999).



\end{thebibliography}
\end{document}